# Autoionization dynamics of He nanodroplets resonantly excited by intense XUV laser pulses

Y. Ovcharenko[1,10,*], A. LaForge[2], B. Langbehn[1], O. Plekan[3], R. Cucini[3], P. Finetti[3], P.O'Keeffe[4], D. Iablonskyi[5], T. Nishiyama[6], K. Ueda[5], P. Piseri[7], M. DiFraia[3,8], R. Richter[3], M.Coreno[3,4], C.Callegari[3], K. C. Prince[3,9], F. Stienkemeier[2,11], T. Möller[1,*] and M. Mudrich[12,*]

[1]Institut für Optik und Atomare Physik, TU Berlin, 10623 Berlin, Germany
[2]Physikalisches Institut, Universität Freiburg, 79104 Freiburg, Germany
[3]Elettra-Sincrotrone Trieste, Basovizza, 34149 Trieste, Italy
[4]ISM-CNR, Area della Ricerca di Roma 1, Monterotondo Scalo, Italy
[5]Institute of Multidisciplinary Research for Advanced Materials, Tohoku University, 980-8577 Sendai, Japan
[6]Division of Physics and Astronomy, Graduate School of Science, Kyoto University, 606-8501 Kyoto, Japan
[7]CIMAINA and Dipartimento di Fisica, Università degli Studi di Milano, 20133 Milano, Italy
[8]Department of Physics, University of Trieste, 34128 Trieste, Italy
[9]IOM-CNR TASC Laboratory, Basovizza, 34149 Trieste, Italy
[10]European XFEL GmbH, 22607 Hamburg, Germany
[11]Freiburg Institute of Advanced Studies (FRIAS), University of Freiburg, 79194 Freiburg, Germany
[12]Department of Physics and Astronomy, Aarhus University, 8000 C Aarhus, Denmark

**Abstract:** The ionization dynamics of helium droplets in a wide size range from 220 to $10^6$ He atoms irradiated with intense femtosecond extreme ultraviolet (XUV) pulses of $10^9 \div 10^{12}$ W/cm$^2$ power density is investigated in detail by photoelectron spectroscopy. Helium droplets are resonantly excited in the photon energy range from ~ 21 eV (corresponding to the atomic 1s2s state) up to the atomic ionization potential (IP) at ~ 25 eV. A complex evolution of the electron spectra as a function of droplet size and XUV intensity is observed, ranging from atomic-like narrow peaks due to binary autoionization, to an unstructured feature characteristic of electron emission from a nanoplasma. The experimental results are analyzed and interpreted with the help of numerical simulations based on rate equations taking into account various processes such as multi-step ionization, interatomic Coulombic decay (ICD), secondary inelastic collisions, desorption of electronically excited atoms, collective autoionization (CAI) and further relaxation processes.

* Corresponding authors: thomas.moeller@physik.tu-berlin.de
yevheniy.ovcharenko@xfel.eu
mudrich@phys.au.dk

# I. INTRODUCTION

The rapid development of short-wavelength free-electron lasers [1, 2, 3] (FELs) during recent decades is stimulating the investigation of the interaction between intense, high-energy light pulses and matter, and indeed it has become a very active field of research in atomic and molecular science [4, 5, 6]. Key questions are related to the ionization dynamics on an atomic level and on a sub-fs time scale, answers to which will help to develop an understanding of ionization processes in more complex systems. In pioneering experimental and theoretical studies, various new phenomena such as absorption enhancement [4, 7] and bleaching [6, 8, 9] as well as modification [10] and suppression [11] of electron emission have been discovered.

A detailed understanding of these mechanisms is of fundamental interest and is particularly important for many future studies with novel light sources which are expected to open new fields in spectroscopy and x-ray imaging, such as recording movies of ultrafast processes and chemical reactions [12]. Depending on the power density, samples can absorb a large number of photons and thus be transformed into a highly excited, non-equilibrium state within femtoseconds. In this context, clusters play an important role. Tuning their size allows us to investigate *intra-atomic* vs. *inter-atomic* mechanisms, and even collective phenomena, thus bridging the gap between molecular and condensed matter physics.

Once a nanoscale object(e.g. a large molecule or cluster) absorbs more than one photon, ultrafast energy exchange between the constituents of the object is expected to crucially impact the relaxation dynamics. Ionization by intense XUV pulses is heavily influenced by the 'complex' electron dynamics, either due to multi-electron collisions with energy exchange [10] or by novel types of autoionization processes related to interatomic Coulombic decay (ICD), as predicted recently [13]. According to that work, clusters resonantly irradiated by intense light pulses with photon energy insufficient to ionize t h e atoms by single photon absorption, efficiently autoionize due to energy exchange between two excited electrons. Neighbouring excited atoms exchange energy resulting in the ionization and emission of the electron from the cluster (Fig. 1a). After autoionization the free electron either ionizes the third excited atom (Fig. 1c) or promotes the excited electron to higher Rydberg states (Fig.1.d). In the case when three excited atoms are in direct contact (Fig.1b), inelastic collisions are expected to be even more efficient and dominant [14].

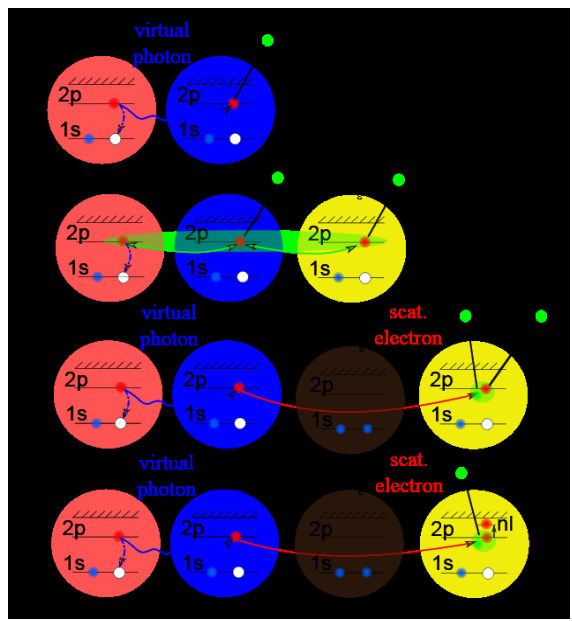

**Figure 1.** Schematic diagram of (a) ICD process [13], (b) CAI type process involving three excited atoms, and (c), (d) CAI with electron scattering on a third neighbouring excited atom [14].

Since this process is based essentially on the sequential absorption of single photons, it is very efficient and thus can easily outrun multi-photon processes [13, 14]. With increasing power density an unusual form of multiply-excited, but rather cold plasma- like state forms which is expected to autoionize on a fs to ps-time scale. The first evidence for such an ionization process in Ne clusters has been reported recently [15, 16, 17].

Subsequent experimental work on He nanodroplets showed a strong enhancement of ionization rates upon resonant excitation of the nanodroplets with respect to direct ionization [14, 18]. This was explained by an ultrafast collective autoionization process (CAI) related to ICD, where several electronically excited atoms are involved. This process was identified through electron spectroscopy - [14], where the dynamics of He nanodroplets resonantly excited to the 1s2p atomic-like states [19] by intense femtosecond XUV pulses was investigated. The electron spectra reveal that in this case, a high-density nanoplasma with a large number of electrons in bound excited states is formed. The novel ionization mechanism is characterized by fast energy exchange and subsequent autoionization of at least three electrons in excited states [14]. In that study, however, many questions still remained open, especially pertaining to the transition from two-body ICD to complex many-body autoionization, as a function of power density and droplet size.

Therefore, precise measurements using simple model systems can add to the fundamental understanding of such decay processes in the condensed phase. Here, we present experiments and simulations with helium (He) nanodroplets which feature (i) an extremely simple electronic structure of the He constituent atoms, (ii) extremely weak interatomic van der Waals interactions, and (iii) a homogeneous, superfluid density distribution which is nearly independent of the droplet size [2, 3].

In the present study, we address the autoionization dynamics, especially the transition from two-body ICD to complex many-body autoionization. We present detailed investigations, showing the dependence of photoelectron spectra with respect to the droplet size and power density. Additionally, a through description of the processes involved based on numerical simulations is also given. The article is organized as follows. In Sec.II we briefly describe the experimental setup. The theoretical model, details of the simulations and a system of rate equations are the subject of Sec. III. Experimental and simulation results are discussed in Sec. IV. In the last section, a summary of the results, a conclusion and an outlook are given.

## II. EXPERIMENTAL PROCEDURE

The experiment was performed at the low density matter (LDM) beam line [20] of the FERMI FEL [3]. FEL pulses with photon energies in the energy range $19 - 40$ eV having a wide range of pulse energies (0.2–30 μJ) were focused by a Kirkpatrick-Baez optical system [21] to a spot size of around 300 μm (FWHM) diameter for photon energies hv below the first ionization potential ($I_P$) of He, and to 20 μm (FWHM) diameter for hv>$I_P$, respectively. The FEL polarization was chosen to be linear and the axis to be perpendicular with respect to a detector axis, while the estimated pulse length is 130 fs (FWHM). Taking into account the estimated transmission of all optical components of the beamline (∼ 38 %), the power density in the interaction region is calculated to be in the $10^9$–$10^{12}$ W/cm$^2$ range. The second order radiation of the FEL beam is in order of few percentages. He nanodroplets with an average number of atoms from 220 up to $10^6$ were produced in a supersonic expansion of He gas at 50 - 80 bar stagnation pressure through the conical nozzle (100 μm diameter, half-opening angle of 45°) cooled to a temperature of 28 − 5 K with a precision of ± 0.1 K. The droplet size is determined based on titration measurements performed separately [22]. The kinetic energy distribution of emitted electrons was measured using a velocity map imaging (VMI) spectrometer. The kinetic energy distributions were reconstructed using a standard Abel inversion method [23], taking into account the calibration curve of the VMI spectrometer. The uncertainty of the energy calibration, ΔE/E, was determined to be less than 4% for electron kinetic energies above 10 eV and less than 12% for energies below 10 eV.

Experimental photoelectron spectra for direct photoionization of He nanodroplets at $h\nu$ =42.8 eV (> $I_p$ =24.59 eV) and various power densities are given in Fig. 2a).

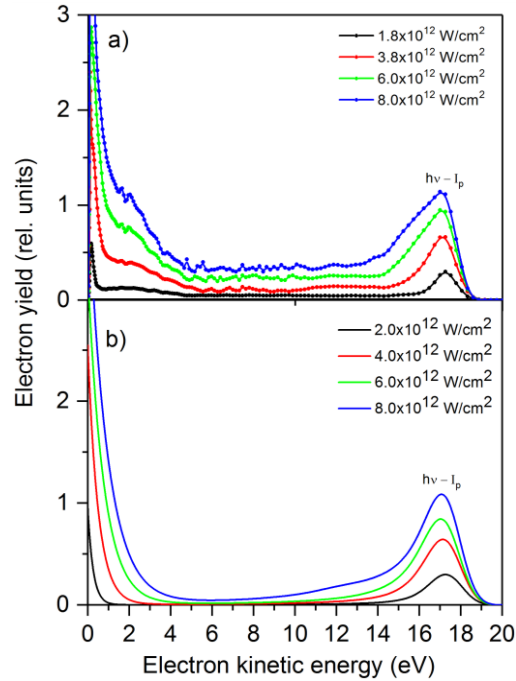

**Figure 2.** Photoelectron spectra of He nanodroplets ($N \approx$ 50 000 atoms) irradiated at 42.8 eV photon energy and different power densities: a) experimental results, b) simulations.

The FEL intensity is varied between I=$1 \times 10^{12}$ − $8 \times 10^{12}$ W/cm$^2$ using a gas cell attenuator [24] and the data are sorted into bins according to the incoming pulse intensities. VMIs averaged for each bin are converted to kinetic energy distributions by inverse Abel transformation. In this work, the inversion of the background subtracted 2D images was performed using the pBasex method introduced by Garcia et al. [23] combined with a Windows type graphical user interface which has extensive image manipulation features [program available from the authors on request.]. For the highest power densities (see Fig. 2a) there are at least two prominent maxima: a first one around 17.0 eV due to direct photoelectron emission; a second one around 0 eV which exponentially decays towards increasing energies, indicative of evaporative electron emission from a nanoplasma [25].

## III. Numerical simulation

The full dynamics of the ICD and CAI processes induced in He nanodroplets by resonant excitation of the 1s → 2p transition by strong EUV laser fields ideally should be simulated using a quantum mechanical description, which unfortunately currently cannot be achieved. Instead, semi-classical approaches are widely used to model interaction processes in clusters [26, 27]. These are based on the description of the atomic ionization processes via suitable rates, whereas the dynamics of the resulting ions and electrons is treated by classical dynamics [28]. Unfortunately, this approach is difficult to apply to resonantly exited He nanodroplets due to the contribution of several processes and their correlations. Therefore, at present, our experimental results can only be treated by numerical simulations, i.e. via Monte Carlo sampling, based on a simplified model given by a system of rate equations of various processes such as multi-step ionization [10, 25], interatomic Coulombic decay [13], secondary inelastic collisions [10, 14] and desorption of electronically excited atoms in He droplets [29], as well as other relaxation processes discussed below.

As has been shown in [14], at a photon energy well above the first $I_P$ of the He nanodroplet, photoelectron spectra can be interpreted by a sequence of direct electron emission events in the developing Coulomb field, which is called multi-step ionization [11, 25]. While the droplet absorbs many photons, electrons are ejected one after another thereby charging up the droplet. As a result, electrons emitted at later stages need to overcome the Coulomb potential created by the charged droplet and thus lose more energy [11, 25]. The escape of electrons is described as an instantaneous process, i.e., the emitted electrons leave the droplet before the next ionization event occurs, thereby neglecting any further energy exchange. Assuming a droplet ionization process as a series of instantaneous electron emission events due to direct photoemission from the developing droplet Coulomb field and accepting that the ionization events are counted only if the single-particle energy of the released electron is positive, the asymptotic kinetic energy of an electron released from the j-th ion is determined by [11, 25]

$$E_j = h\nu - I_p - \frac{e^2}{4\pi\varepsilon_o}\sum_{i \neq j}\frac{q_i}{r_{ij}}, \qquad (1)$$

where $h\nu$ is a photon energy, $I_p$ is the ionization potential and $i$ runs over all other ions with charge state $q_i$ (integer number) and distance between electron and ions, $r_{ij}$. The last term in Eq. (1) describes the Coulomb downshift due to the previously generated ions with charge states $q_i$ (in our case $q_i = 1$) at positions $r_i$. In this way, Monte Carlo simulations of the photoelectron spectra can be performed, where only direct ionization events, i.e., electrons with kinetic energy $E_k > 0$, are considered. At sufficiently high power density, the emission process stops since the electrons cannot escape from the high Coulomb potential of the droplet and the electron emission becomes frustrated [11].

Monte-Carlo simulations of multi-step photoionization of He nanodroplets at hν=42.8 eV photon energy (peak structure between 2 and 14 eV) in combination with the characteristic tail due to evaporative electron emission (exponentially falling distribution between 0 and 2 eV) are shown in Fig. 2b). Since experimental and simulated results are in good agreement, we conclude that He nanodroplets irradiated at hν> $I_p$ follow similar photoionization dynamics as small heavier rare-gas clusters, which feature a characteristic plateau in the photoelectron spectrum [11, 25]. Electrons released by autoionization should be affected in the same way by the developing Coulomb field as photoelectrons. Therefore, the effect of multi-step ionization – broadening of photolines and the formation of a plateau – is also taken into account in our simulations for hν < $I_p$.

In the case of hν=21.5 eV, autoionization can only be triggered by efficient interatomic electronic decay processes like ICD [13] or by two-photon ionization processes. Considering that the photoionization cross section of He excited states ($2^1$S, $2^3$S, $2^1$P) is below 0.05 Mb [30] and the power density of the FEL beam is below $10^{11} W/cm^2$, two-photon ionization processes are not expected to significantly contribute to the ionization signals and are therefore neglected in our model. As shown in [18], the fraction of He nanodroplets ionized by sequential two-photon ionization at hν=21.6 eV and I=$10^{13}$ W/cm² is expected to be 0.6 % while non-sequential two-photon ionization is not observed. Stimulated emission is neglected due to efficient depletion of He excited states via ICD. Therefore, in our model for multiply excited He nanodroplets, we assume ICD to be the only significant ionization process which we refer to as multi-step ICD. Since the ICD mechanism is expected to be very efficient [13], one assumes that every pair of He droplet excited states will undergo ICD producing one ionized and one neutral atom, respectively.

Furthermore, fast electronic relaxation [31] of the droplet excited states to low lying $2^1$S, $2^3$S, $2^1$P atomic excited states has been observed in our experimental results (see Fig. 3). Here, the electron spectra of He droplets with N=2800 atoms irradiated at different photon energies are presented. As seen, the position of the main peak at 16.6 eV (feature "a"), which corresponds to ICD of pairs of 1s2s $^3$S atomic states, is independent of the photon energy, while the position of the features "b", "c" and "d" due to the second harmonic radiation is changed. Thus, when He droplet is irradiated at hν=21.5 eV (dotted red line, –·–·–), the feature "c" on the right hand shoulder at 18.43 eV electron energy matches the maximum of atomic electron spectrum irradiated at hν=43.0 eV (dashed black line, second harmonic of 21.5 eV). In the case of hν = 21.02 eV (dotted blue line, –·–·–) and hν = 22.6eV (dotted green line, –·–·–), lower (feature "b" at 17.5 eV) and higher shifts by ∼ 1 eV (feature "d" at 20.8 eV) with respect to maximum "c" are observed. For hν = 22.6 eV, the peak around 16.6 eV is smaller than the one around 20.8 eV for the dotted green line ( –·–·– ), since the direct photoionization by the second harmonic is more efficient than ICD due to the low excitation cross section.

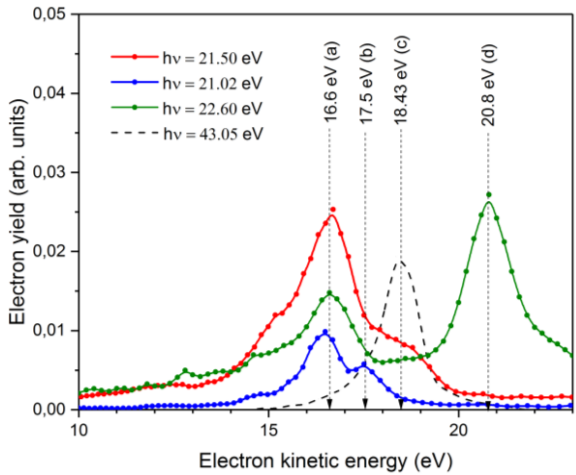
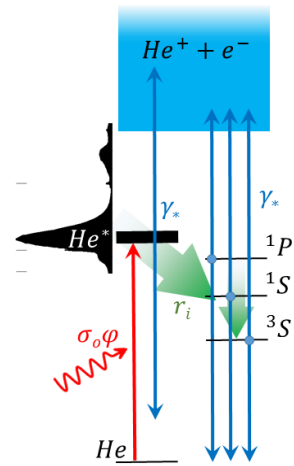

**Figure 3.** Electron spectra of He nanodroplets ($N$ = 2,800 atoms) irradiated at 21.5 eV (dash dotted red line, –·–·–), 21.02 eV (dash dotted blue line, –·–·–) and 22.6 eV (dash dotted green line, –·–·–) FEL photon energies with ∼1.5x10$^{10}$ W/cm$^2$ power density. The dashed black (––) line with maximum (c) at 18.43 eV represents the photoelectron spectrum of an atomic beam irradiated at the photon energy of 43.05 eV. Feature (a) corresponds to ICD of a pair of atoms in 1s2s $^3$S states [32] within a droplet; features (b) and (d) show lower and higher shifts of ∼1 eV with respect to maximum (c).

**Figure 4.** Schematic representation of possible decay channels of excited He nanodroplet at 21.5 eV photon energy. The He droplet photoabsorption spectrum is indicated in black [19]. After excitation of a He droplet at 21.5 eV the droplet most probably relaxes to low lying 2s and 2p atomic excited states (green arrows, →), followed by ICD (blue arrows, →).

The fast electronic relaxation process of He droplet excited states decaying to low lying 2$^1$S, 2$^3$S, 2$^1$P atomic excited states has recently also been observed by measuring the decay time of band-excited states in pump-probe experiments [31, 33]. It is shown that the band-excited state at hν=21.5 eV in He nanodroplets consisting of about 40,000 atoms decays to the atomic 1s2p and 1s2s states within few tens of fs. Subsequently, 1s2p ($^1$P$_1$) → 1s2s ($^1$S$_o$) relaxation takes place with a characteristic decay time of ∼400 fs, followed by slow 1s2p ($^1$S$_o$) → 1s2s ($^3$S$_1$) relaxation (ps time scale) [31, 33]. The relaxation time may slightly depend on droplet size, since the average atomic density is size-dependent. Various important relaxation processes in He nanodroplet are schematically depicted in Figure 4. At the beginning, 2p droplet excited states $N^*$ undergo fast relaxation to the atomic 1s2p $^1$P, 1s2s $^1$S and 1s2s $^3$S states with the relaxation rate constants $r_i$ followed by ICD between pairs of these states at the decay rate constant $\gamma_*$. Based on our experimental finding (see subsection IV. A), we assume that ICD can only occur between 1s2s $^3$S, 1s2s $^1$S and 1s2p $^1$P pairs of identical excited states within a droplet. Even more, this assumption is sufficient to describe our results.

Based on the processes mentioned above as well as the proposed way of computing the number of excited and ionized states in [13] by solving a system of rate equations, we model the evolution of excited He droplets by the following system of rate equations

$$\dot{N}(t) = -\sigma_o \cdot \varphi(t) \cdot N(t) + \sum_i r_i \cdot \gamma_* \cdot N^*(t)^2 \qquad (2a)$$

$$\dot{N}^*(t) = \sigma_o \cdot \varphi(t) \cdot N(t) - 2 \cdot \sum_i r_i \cdot \gamma_* \cdot N^*(t)^2, \qquad (2b)$$

$$\dot{N}^+(t) = \sum_i r_i \cdot \gamma_* \cdot N^*(t)^2. \qquad (2c)$$

Here, $\sum_i r_i \cdot \gamma_* \cdot N^*(t)^2$ is the sum of ICD decay rates over $i$ excited states, i.e., $i$ runs over 1s2s $^3$S, 1s2s $^1$S and 1s2p $^1$P atomic excited states. Equation (2a) corresponds to the time evolution of the neutral ground states, where $N(t)$ is the number of neutral atoms in the droplet as a function of time $t$, $\sigma_o$ is the absorption cross section, and $\varphi(t)$ denotes the photon flux which contains the information about the temporal profile of the pulse. Equation (2b) denotes the time evolution of excited states, where $N^*(t)$ is the number of excited atoms inside the droplet as a function of time $t$. Equation (2c) describes the time evolution of ionized atoms $N^+(t)$ as a function of time $t$.

To simulate the electron spectra, we solve the rate equation model numerically using the Monte-Carlo method. We start the simulation by determining the number of droplet excited states over the whole FEL pulse with a simulation time step of 1 fs based on the FEL power density, FEL pulse width and He droplet excitation cross-section per atom $\sigma_o$ [14, 32]. Assuming fixed positions of excited states in space, we randomly distribute He excited states over the nanodroplet. For every time step of the simulation we evaluate the probability of undergoing a discrete relaxation or ICD ionization process. Final electron spectra are obtained by summing over the modelled multi-step ionization processes [25] of individual ICD components (based on formula (1)):

$$E_J^{ICD} = 2h\nu - I_p - \frac{e^2}{4\pi\varepsilon_o} \sum_{i \neq j} \frac{q_i}{r_{ij}}. \qquad (3)$$

For large He nanodroplets, where high order ICD/CAI processes contribute, the system of rate equations changes correspondingly (see explanation below). Finally, the obtained electron spectra are convoluted by the energy resolution of the VMI spectrometer (see Table 1). Intensity-averaging over the volume of the focus is not taken into account in our simulation. All parameters relevant for the numerical simulation are given in Tab. 1.

| | |
|---|---|
| Excitation photon energy, hν | 21.5 eV |
| FEL pulse width (FWHM), $\Delta\tau$ | 130 fs |
| Spectrometer energy resolution , $\Delta E/E$ | 4 % |
| Gaussian profile of the FEL beam (FWHM) | 300 μm |
| He droplet excitation cross-section per atom [14, 32], $\sigma_o$ | 50 Mbarn |
| Electron-impact 2p ionization cross-section [34], $\sigma_{2p\rightarrow inf.}$ | 1200 Mbarn |
| Electron-impact 2p→nl excitation cross-section [34], $\sum_{n=3}^{5} \sigma_{2p\rightarrow nl}$ | 2800 Mbarn |
| Droplet relaxation time to 1s2p and 1s2s atomic excited states [31, 33] | < 2.5 ps |
| 1s2p ($^1P_1$) → 1s2s ($^1S_0$) relaxation time [31, 33] | < 5 ps |
| Radiative decay time of 1s2s ($^1S_0$) and 1s2s ($^3S_1$) atomic excited states | > 100ns |
| ICD decay time, assumption | 500 fs |
| He desorption time [35] | 15 ps |
| Ionization potential, $Ip$ | 24.59 eV |
| Internuclear separation [36] | 3.36 A |
| Simulation time step | 1 fs |
| Number of iterations per step point | 10000 |
| Time interval with respect to the FEL pulse | from -500 fs to +1500fs |

Table 1: Numerical factors used in the simulation of the ionization dynamics of 1s2p $^1P_1$ excited He droplets.

As an example, the simulated ionization dynamics of a small size He nanodroplet (250 atoms) resonantly irradiated at hν = 21.5 eV by a 130-fs FEL pulse at I=2.4x10$^{10}$ W/cm$^2$ is shown in Figure 5. When the FEL beam interacts with He nanodroplets, a large numbers of $2p$-excited states ($N^*$) are formed resulting in "quasi"-free electrons in the droplets due to ICD. As has been shown in [10] and [14], inelastic electron collisions in this case are very efficient. Therefore, the number of excited He atoms $N_{excit.}(t)$ (see Fig.1(d)) and the number of ionized atoms $N_{ioniz.}(t)$ (see Fig.1(c)) by inelastic electron collisions as a function of time $t$ is given by

$$N_{inel.}(t) = N_{excit.}(t) + N_{ioniz.}(t), \qquad (4a)$$

where

$$N_{excit.}(t) = \sum_{n=3}^{5} \sigma_{2p\rightarrow nl} \cdot N^+(t) \cdot N^*(t) \qquad (4b)$$

$$N_{ioniz.}(t) = \sigma_{2p\rightarrow inf} \cdot N^+(t) \cdot N^*(t). \qquad (4c)$$

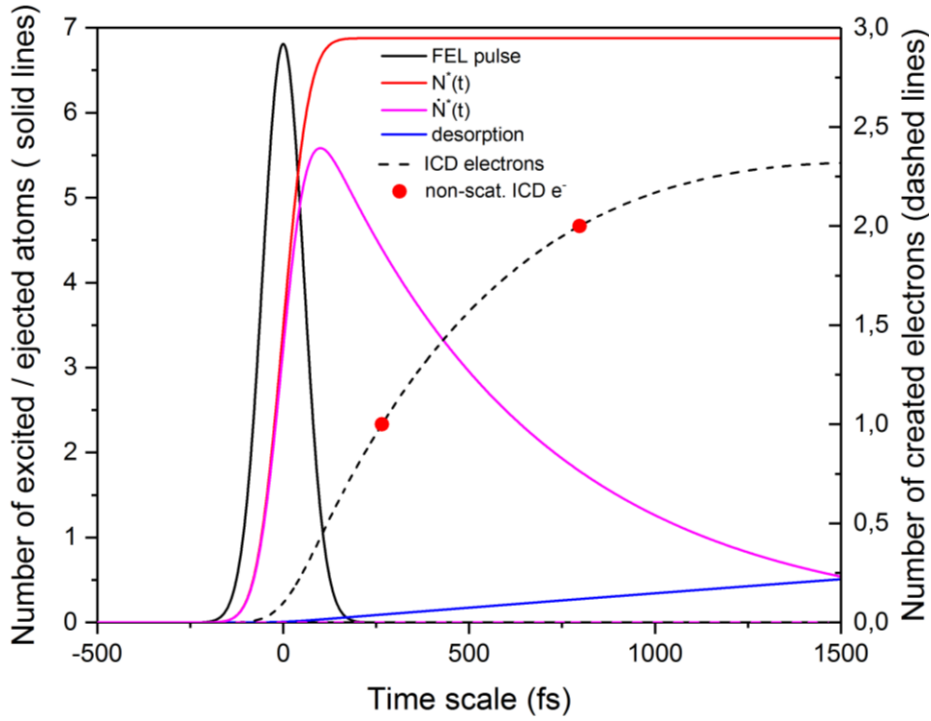

**Figure 5.** Simulated ionization dynamics of $1s2p\ ^1P_1$ excited states within $He_{250}$ nanodroplets irradiated by 130-fs FEL pulses (solid black line, −) and at $2.4 \times 10^{10}$ W/cm$^2$ power density. The total number of excited atoms $N^*(t)$ is shown by the solid red (−) line. The evolution of excited states $\dot{N}^*(t)$ due to ICD and desorption (solid blue line, −) of the excited atoms is shown by the solid magenta (−) line. The total number of created ICD electrons and the discrete number of non-scattered ICD electrons (non.-scat. ICD e⁻) are shown as a dashed black line (--) and as red dots (●), respectively. For this droplet size and power density the inelastic electron scattering process is negligible.

Here $\sigma_{2p \to nl}$ and $\sigma_{2p \to inf}$ are the electron-impact excitation and electron-impact ionization cross sections of the He atom from the 2p excited state to high lying *nl* levels and to the continuum, respectively. Equation (3b) contains the sum of $\sigma_{2p \to nl}$ electron-impact excitation cross sections, where the principal quantum number *n* runs over the states whose contributions to the total cross section is largest.

Another important relaxation process, which also has to be considered, is the highly efficient desorption of electronically excited atoms and molecules from He droplets [29, 37]. As has been explained in [29], excited atoms He* and molecules (excimers) He₂* tend to form bubble states around them due to Pauli repulsion of the outer electron and the surrounding He. Owing to the superfluid state of the He nanodroplets, the bubble states freely move to the surface where the He* and He₂* are ejected into vacuum. Since the electronically excited molecules desorbing from the droplet are far from being thermalized, the coupling between the electronically excited molecule and the droplet is very weak. Therefore we assume that the excited species leave the droplet in a very short time [37]. This effect is taken into account by

$$N_{des.}(t) = k \cdot N^*(t),\qquad(5)$$

where *k* is the desorption rate constant.

In this way, electron spectra of He nanodroplets where inelastic electron collisions start to play a role have been modelled using the rate equation system

$$\dot{N}(t) = -\sigma_o \cdot \varphi(t) \cdot N(t) + \sum_i r_i \cdot \gamma_* \cdot N^*(t)^2 \tag{6a}$$

$$\dot{N}^*(t) = \sigma_o \cdot \varphi(t) \cdot N(t) - 2 \cdot \sum_i r_i \cdot \gamma_* \cdot N^*(t)^2 - N_{des.}(t) - N_{inel.}(t), \tag{6b}$$

$$\dot{N}^+(t) = \sum_i r_i \cdot \gamma_* \cdot N^*(t)^2 + N_{ioniz.}(t). \tag{6c}$$

As an example, the simulated ionization dynamics of He nanodroplets (50 000 atoms) resonantly irradiated at hv =21.5 eV by 130-fs FEL pulses at I=2.6x10$^{10}$ W/cm$^2$ is shown in Figure 6.

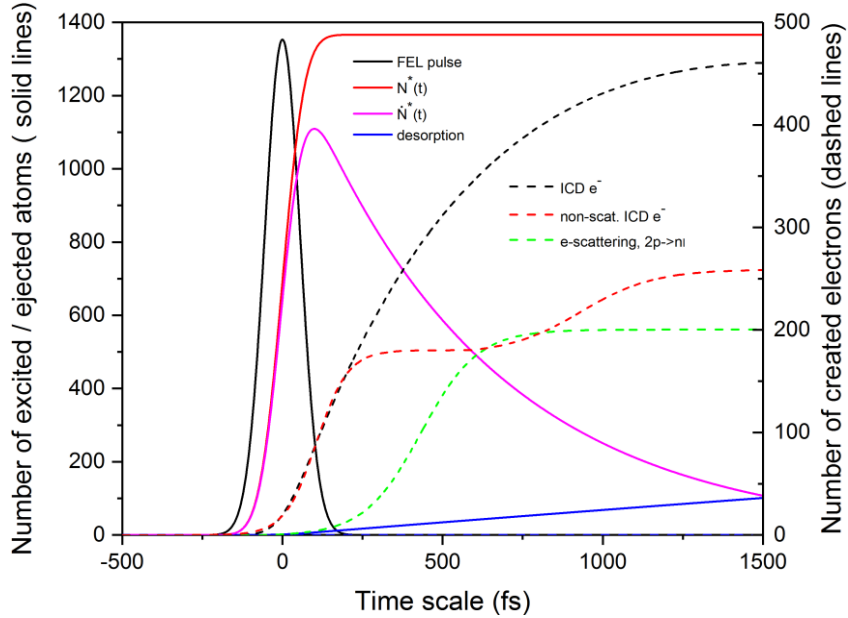

**Figure 6.** Simulated ionization dynamics of 1s2p $^1P_1$ excited states within He$_{50\ 000}$ nanodroplets by 130-fs FEL pulses (solid black line, −) at 2.4x10$^{10}$ W/cm$^2$ power density. The total number of exited atoms $N^*(t)$ is shown by the solid red (−) line. The evolution of excited states $\dot{N}^*(t)$ due to both ICD decay and desorption (solid blue line, −) of the excited atoms is given by the solid magenta (−) line. The total number of created ICD electrons and the discrete number of non-scattered ICD electrons (non-scat. ICD e$^-$) are shown by the dashed black (--) and dashed red (--) lines, respectively. Inelastic electron-impact excitation of 2p excited states to high-lying $nl$ levels (e-scattering, 2p->nl) is shown by the dashed green (--) line, respectively.

In the case of large He nanodroplets, (see IV. RESULTS, subsection C), the ICD electron can scatter several times within a droplet thereby losing part or all of its energy in each collision. If more than two particles are involved in the process, the collective autoionization process (CAI) [14, 18] takes place. The corresponding term in the rate equation is

$$N_{CAI}(t) = N_{excit.}^{CAI}(t) + N_{ioniz.}^{CAI}(t), \tag{7a}$$

where

$$N_{excit.}^{CAI}(t) = \sum_{n=3}^5 \sigma_{2p \to nl} \cdot N_{excit.}(t) \cdot N^*(t) \tag{7b}$$

$$N_{ioniz.}^{CAI}(t) = \sigma_{2p \to inf} \cdot N_{ioniz.}(t) \cdot N^*(t). \qquad (7c)$$

Thus, the system of rate equations for modelling the electron spectra of large He nanodroplets takes the form

$$\dot{N}(t) = -\sigma_o \cdot \varphi(t) \cdot N(t) + \sum_i r_i \cdot \gamma_* \cdot N^*(t)^2 \qquad (8a)$$

$$\dot{N}^*(t) = \sigma_o \cdot \varphi(t) \cdot N(t) - 2 \cdot \sum_i r_i \cdot \gamma_* \cdot N^*(t)^2 - N_{des.}(t) - N_{inel.}(t) - N_{CAI}(t), \qquad (8b)$$

$$\dot{N}^+(t) = \sum_i r_i \cdot \gamma_* \cdot N^*(t)^2 + N_{ioniz.}(t) + N_{ioniz.}^{CAI}(t). \qquad (8c)$$

As an example, the simulated ionization dynamics of He nanodroplets (50 000 atoms) resonantly irradiated at hν =21.5 eV by 130-fs FEL pulses at I=7.5x10$^{10}$ W/cm$^2$ is shown in Fig. 7.

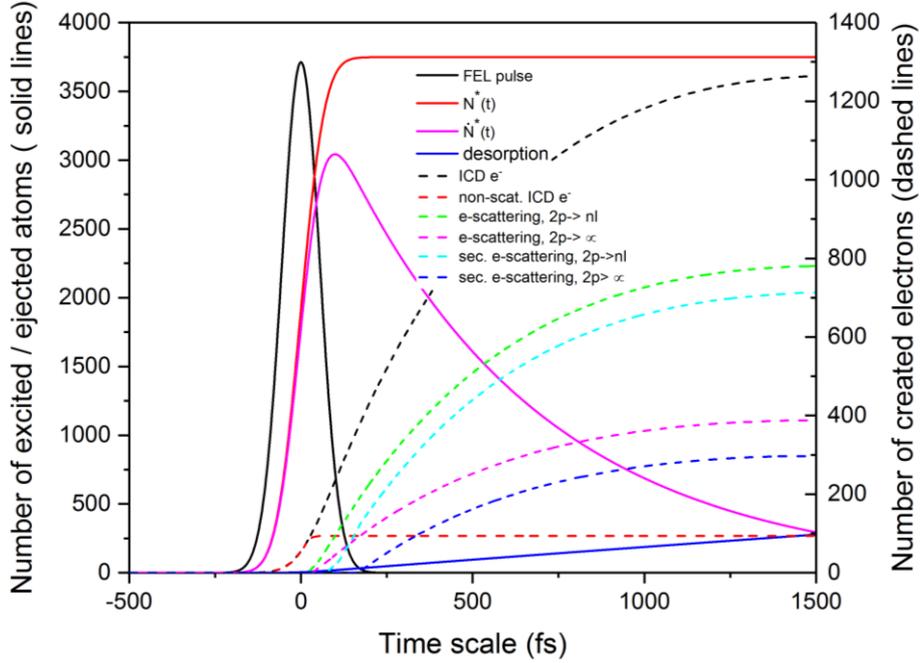

**Figure 7.** Simulated ionization dynamics of 1s2p $^1P_1$ excited states within He$_{50\,000}$ nanodroplets by130-fs FEL pulses (solid black line, −) at I=7.5x10$^{10}$ W/cm$^2$. The total number of excited atoms is shown by the solid red (−) line. The solid blue line (−) shows the desorption process. The evolution of excited states is given by the solid magenta (−) line, which represents the number of He excited atoms after both ICD decay and desorption of He excited atoms. Total number of created ICD electrons (ICD e$^-$) and discrete number of non-scattered ICD electrons (non-scat. ICD e$^-$) are shown by the dashed black (--) and dashed red (--) lines, respectively. Inelastic electron-impact excitations of the 2p excited state to high-lying *nl* levels (e-scattering, 2p->nl) is given by the dashed green (--) line. Inelastic electron-impact excitations of the 2p excited state to the continuum (e-scattering, 2p->∞) is given by the dashed magenta (--) lines. Inelastic electron-impact excitations of the 2p excited state to high-lying *nl* levels in the secondary scattering process (sec. e-scattering, 2p->nl) is given by the dashed cyan (--) line. Inelastic electron-impact excitations of the 2p excited state to the continuum in the secondary scattering process (sec. e-scattering, 2p->∞) is given by the dashed blue (--) line.

## IV. RESULTS

In the following we present both experimental and numerical results for the resonant excitation of He nanodroplets irradiated at hν =21.5eV by 130 fs pulses in the $10^{10}$ – $10^{11}$ W/cm$^2$ power density range. To obtain sufficient statistics, the experimental results have been averaged over 3000 shots and ensemble averaging over $10^5$ simulations is performed.

### A. ICD in small size He droplets

To provide a basis for the forthcoming discussions, we start our analysis from the simplest case, i.e., from the electron spectra of small He nanodroplets (~220 atoms/droplet) resonantly irradiated at hν =21.5 eV and relatively low power densities. As has been discussed in the previous section, at these experimental conditions, electrons can only leave the He droplets due to ICD,

$$A^* \cdots A^* \rightarrow A \cdots A^+ + e^-, \qquad (8)$$

as predicted in Ref. [13]. The electron spectra of He droplets irradiated at hν=21.5 eV and I=2.4x10$^{10}$ W/cm$^2$ are simulated by modelling the time evolution of the excited He nanodroplet (see Fig. 5, section III) up to 1.5 ps.

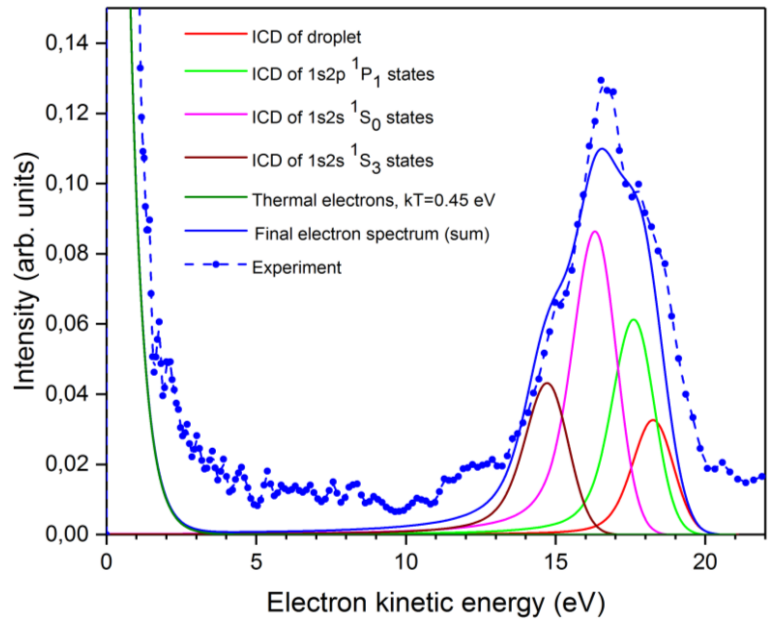

**Figure 8.** Simulated electron spectrum of small He nanodroplets ($N$ = 250 atoms) irradiated at hν=21.5 eV and I=2.4x10$^{10}$ W/cm$^2$ (solid blue (–) line). Electron spectra due to ICD of droplet excited states, atomic 1s2p $^1P_1$, 1s2s ($^1S_0$) and 1s2s ($^3S_1$) states are presented by solid red (–), solid green (–), solid magenta (–) and solid brown (–) lines, respectively. The distribution of thermally evaporated electrons with a temperature of 0.45 eV/$k_B$ is represented by the solid dark green (–) line. The sum of all spectra is shown by the solid blue (–) line. The experimental result is given by the dash-dotted blue (–·–) line.

Detailed information about how the final electron spectrum is formed can be inferred from Fig. 8, where all related processes are shown. As seen, the decay path through the 1s2s $^1S_0$ state gives the largest contribution to the final electron spectrum which can be explained by the fast relaxation of 1s2p ($^1P_1$) state to 1s2s ($^1S_0$) state and subsequent slow 1s2s ($^1S_o$) → 1s2s ($^3S_1$) relaxation (see section III). The sum of electron spectra from different excited states is shown by the solid blue (–) line in Fig.8. The dash-dotted blue (–·–) line shows the experimental result at I=2.4x10$^{10}$ W/cm$^2$. Good agreement between experimental data and numerical simulations is achieved when assuming an ICD time constant of 500 fs for all excited states. This value matches the results in [31, 33]. Furthermore, as it will be shown in subsections B and C, this value assures best agreement between our model and the experimental data in the whole range of droplet sizes and power densities.

The low kinetic energy component of the electron distributions is attributed to thermal evaporation of electrons out of the nanoplasma induced by collisional equilibration of quasi-free electrons [11, 27]. This component is implemented in the simulated electron spectrum by adding a decaying fit function obtained from a polynomial fit of the experimental data (see Fig.9(a)), from which we infer an electron temperature of about 0.45 eV / $k_B$.

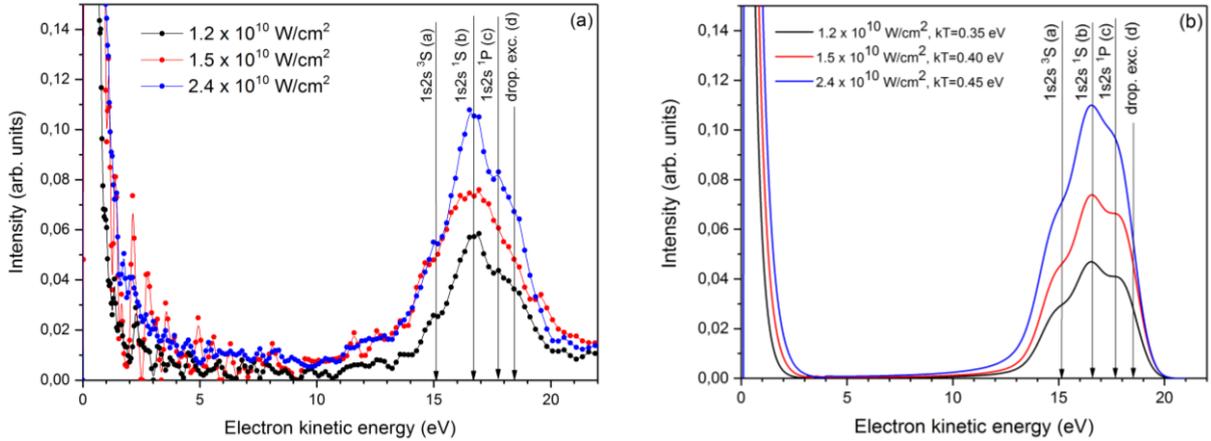

**Figure 9.** Electron spectra of small size He droplets irradiated at hν=21.5 eV and low power densities: a) Experimental results for a He droplet size of 220±60 atoms; b) numerical simulations for a droplet size of 250 atoms. Maxima (a), (b) and (c) correspond to ICD of pairs of 1s2s $^3S$, 1s2s $^1S$ and 1s2p $^1P$ atomic states [32] within a droplet, respectively. Feature (d) is caused by ICD of droplet excited states and direct photoionization of the He atomic beam (for details see text).

As seen from the experimental results (Fig. 9(a)), the electron spectra are rather complicated and do not show a single line as predicted by the pure ICD model [13]. Instead, there are several overlapping peaks which can be assigned to ICD of pairs of 1s2s $^3S$, 1s2s $^1S$ and 1s2p $^1P$ atomic states [32] within a droplet, labelled as "a", "b" and "c", respectively.

(a)  He$^*$(1s2s $^3S$)··· He$^*$(1s2s $^3S$) → He (1s$^2$) ··· He$^+$ (1s) + e$_1$ (15.08 eV)

(b)  He$^*$(1s2s $^1$S)··· He$^*$(1s2s $^1$S) → He (1s$^2$) ··· He$^+$ (1s) + e$_2$ (16.68 eV)          (9)

(c)  He$^*$(1s2p $^1$P)··· He$^*$(1s2p $^1$P) → He (1s$^2$) ··· He$^+$ (1s) + e$_3$ (17.88 eV)

The kinetic energy of the ICD electron is given by E$_e$= 2×E (He*) − E$_i$, where E (He*) is the energy of the excited level of He* and E$_i$ is the ionization potential of He atoms. The influence of the He droplet on the energetics of the ICD is neglected owing to the weak coupling of the excited and ionized atoms to the He droplet surface. The occurrence of electrons from ICD of low-lying atomic excited states gives clear evidence for fast (~100 fs) electronic relaxation (see details in Sec. III). I.e., resonantly excited He nanodroplets relax by droplet-induced transitions to low-lying atomic excited states [37, 38, 39, 40, 41, 42, 35] and then ICD takes place. This is in good agreement with recent experimental results on Ne droplets [43]. Feature "d" at 18.43 eV is caused by ICD of droplet excited states as well as by direct photoionization of the He atomic beam (co-propagates with the droplet beam) by a few percent of the FEL second harmonic radiation.

Taking  into account both ionization pathways and the decay times discussed above as well as multi-step ICD electron emission (see Sec. III), we simulate the electron spectra of resonantly excited He nanodroplets at 21.5 eV (see Fig. 9(b)) for different power densities. As seen, the experimental and simulated results are in good agreement.

## B. ICD in medium-sized He nanodroplets

Experimental electron spectra of medium-sized He nanodroplets (~1000 atoms) irradiated at hν =21.5 eV are shown in Figure 10 (a).

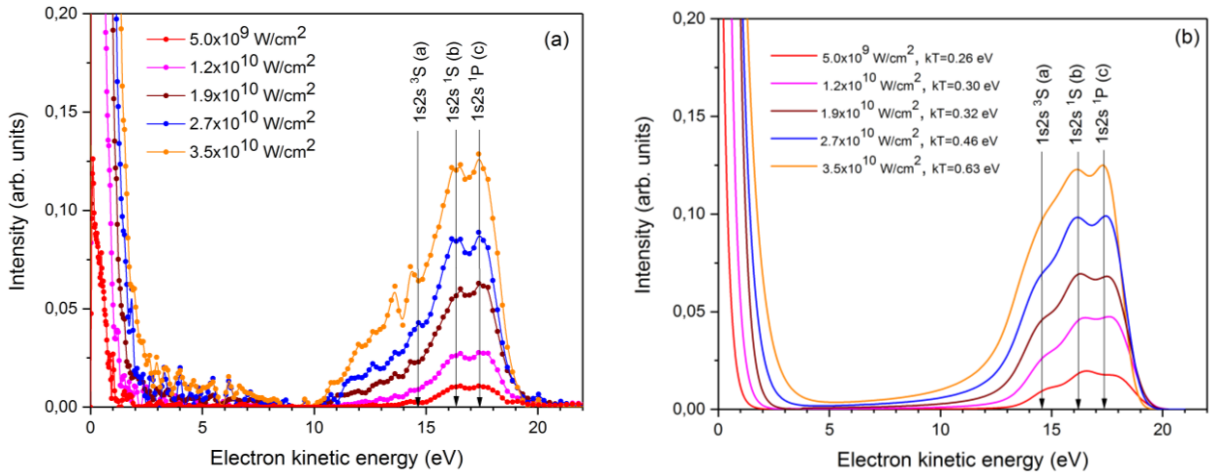

**Figure 10.** Electron spectra of medium sized He droplets irradiated at hν=21.5 eV and low power densities: a) Experimental results for a He droplet size of 1000±300 atoms; b) numerical simulations for a He droplet size of 1000 atoms. Maxima (a), (b) and (c) correspond to ICD of pairs of 1s2s $^3$S, 1s2s $^1$S and 1s2p $^1$P atomic states [32] within a droplet, respectively.

As seen, already under these conditions the structure of the electron spectrum becomes more complicated. Despite the fact that the features around 14.6 eV, 16.3 eV and 17.4 eV are less pronounced than for the small He droplets, they are assigned to ICD of pairs of 1s2s $^3$S, 1s2s $^1$S and 1s2p $^1$P atomic states [32] within a droplet, correspondingly. The same model as for the small He droplets (see Sec. III) has been used for simulating the electron spectra for medium sized He droplets (see Fig. 10(b)).

Based on our numerical simulations and especially on the fact that the same model provides good agreement between experimental and simulated results for two different droplet sizes, we conclude that the decay pathways for He$_{1000}$ nanodroplets are essentially the same as for He$_{250}$ nanodroplets. The broadening of electron spectra is mainly caused by an increasing number of unscattered electrons, which then play a role in the multi-step ICD electron emission.

Increasing the number of atoms within the He nanodroplet up to 2 500 atoms leads to broadening of the spectral features (see Fig.11) due to the creation of a larger number of unscattered ICD electrons. As a consequence, the spectral components from different states merge to one broad peak, which shifts toward lower kinetic energies when increasing the power density (see Fig.11).

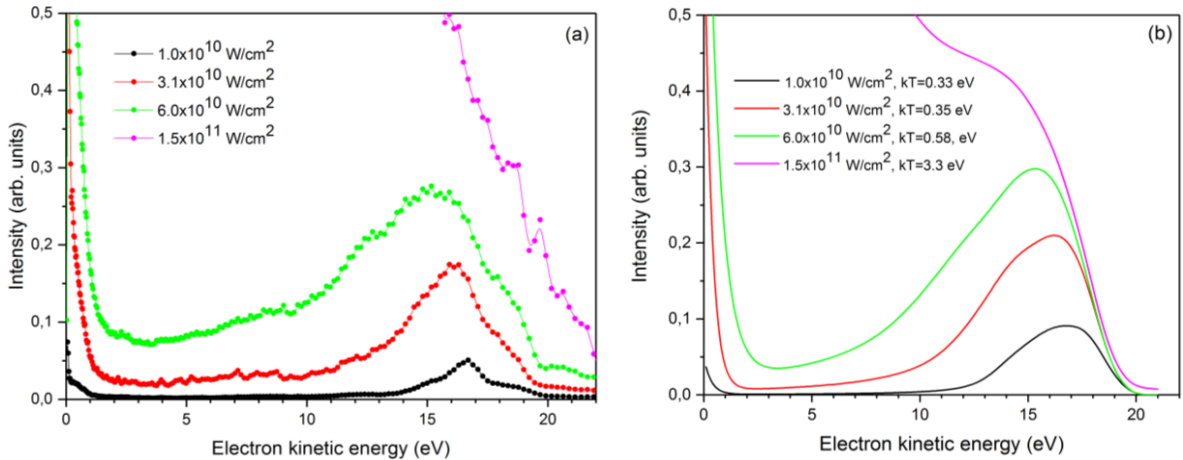

**Figure 11.** Electron spectra of medium-size He droplets irradiated at hv=21.5 eV and low power densities: a) Experimental results for He droplets composed of 2 500±500 atoms; b) numerical simulations for He droplet size of 2 500 atoms.

In contrast to the small He nanodroplets, it becomes difficult to unambiguously assign the various decay channels and their contributions to the electron spectra. Nevertheless, our model still nicely reproduces the experimental electron spectra for resonantly excited He nanodroplets of 25 00 atoms. There are some deviations between experimental and simulated main peak widths for the low power densities (I~ 10$^{10}$ W/cm$^2$), but the peak positions remain the same for experimental and simulated cases. The broadening of the maxima in the numerical simulations can be explained by the fixed values of the decay rates in the simulation. In reality, these values can slightly vary depending on both droplet size and power density.

Our simulation shows that even at I=1.5x10$^{11}$ W/cm$^2$, inelastic electron scattering is negligible for medium-sized He nanodroplets. Nevertheless, the structure of the electron spectra is drastically blurred by the large number of quasi-free electrons in collisional equilibrium [11, 25], which leads to thermal electron emission [28]. Moreover, it relates to a clear transition from ICD autoionization to CAI where at least three excited atoms are in direct contact (see Fig. 1b). The corresponding electron spectral component, from which we infer an electron temperature in the range of 0.3 - 3.3 eV eV, is included into the simulation results (see Fig. 11b).

## C. ICD in large He nanodroplets

Resonant excitation of large He nanodroplets, i.e., in the case when the number of atoms per droplet exceeds 50000 atoms, leads to additional structures and broadening of the electron spectra (see Fig.12). As mentioned in section III, ICD electrons can scatter several times within a droplet, thereby losing part of their energy. Additionally, when more than two electronically excited atoms are involved, a transition from ICD to CAI [14, 18] takes places.

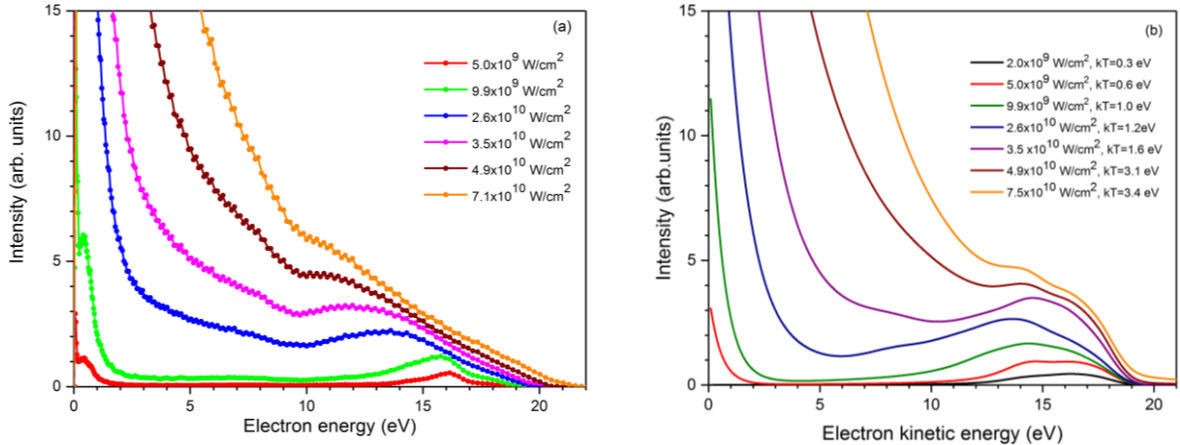

**Figure 12.** Electron spectra of large He droplets irradiated at hν=21.5 eV and low power densities: a) Experimental results for He droplet size of 50 000±1 000 atoms; b) numerical simulations for He droplet size of 50 000 atoms.

For example, already at I=2.6x10$^{10}$ W/cm$^2$ (see Fig. 12(a), dotted blue (–·–) line) a sufficiently large number of ICD electrons can scatter on 2p excited atoms within He$_{50\,000}$ nanodroplets (see Fig. 6, section III), and further excite the atoms to high-lying *nl* excited states. The ionization dynamics of atomic 1s2p $^1$P$_1$ excited states within He$_{50\,000}$ nanodroplets at the 130-fs FEL pulses and I=2.4x10$^{10}$ W/cm$^2$ is shown in Figure 6 (see section III.). A similar behavior is seen for 1s2s $^3$S, 1s2s $^1$S atomic excited states and droplet excited states. As a result, ICD-created electrons from all atomic excited states lose energy due to scattering leading to additional structures in the electron spectra. Additionally, the electron spectra of the individual components are strongly broadened due to multi-step ICD electron emission [11, 25].

Further increase of the power density in He$_{50\,000}$ nanodroplets even leads to multiple scattering events. Thus, the electrons created by ICD scatter on 2p excited atoms, transfering them to higher-lying *nl* excited states or to the continuum, see the dashed green (--) and dashed magenta (--) lines in Fig. 7, respectively. In this way, the ICD electron loses its kinetic energy, depending on the scattering pathway. Since the number of 2p-excited states is still large enough, the same ICD electron can further scatter leading to secondary scattering processes, represented by the dashed cyan (--) and dashed blue (--) lines, respectively. In this way, the kinetic energy of ICD electrons is reduced.

As seen in Fig. 12, there is a good agreement between experimental and simulated results. Our experimental data clearly show that CAI [14] and thermal electron emission provide the main contributions to the electron spectra. At I=7.5x10$^{10}$ W/cm$^2$, the primary contribution to the electron spectra stems from 'thermal' electrons (see Fig.13).

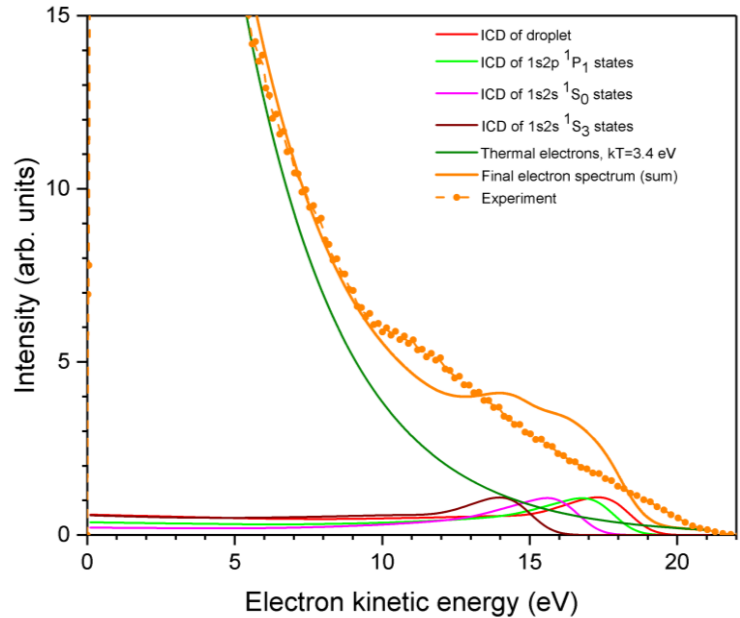

**Figure 13.** Simulated electron spectra of large He nanodroplets ($N = 50\,000$ atoms) irradiated at hv=21.5 eV and I=7.5x10$^{10}$ W/cm$^2$ (solid orange (–) line in Figure 12b). Electron spectra from ICD of droplet exited states and atomic 1s2p $^1$P$_1$, 1s2s ($^1$S$_0$) and 1s2s ($^3$S$_1$) states are presented by solid red (–), solid green (–), solid magenta (–) and solid brown (–) lines, respectively. Thermal electron evaporation at the electron temperature of 3.4 eV/k$_B$ is given by the solid dark green (–) line. The sum of all spectra is shown by the solid orange (–) line. The experimental result from Figure 12a) is given by the dotted orange (----) line.

This means that a large fraction of created electrons in a broad energy range at some point cannot overcome the Coulomb barrier, which leads to the frustration of multi-step ICD electron emission [25]. Thereby the quasi-free electrons are trapped in He nanodroplets leading to the formation of a cold plasma and to the evaporative emission of thermal electrons. Similar to the case of small He droplets, the contribution of the thermal electrons is taken into account by a polynomial fit of the experimental results with an electron temperature of 3.4 eV / $k_B$.

When increasing the number of atoms in He nanodroplets up to 1 000 000, the general behavior of the electron spectra does not change much (see Fig.14). Here, thermal electron emission plays the main role even at lower power densities. Additionally, as shown in [14], a high-density plasma with broad electron features can be formed due to the fact that more than 3 or 4 excited atoms are in direct contact (see Fig. 1b), leading to the formation of a "continuous network", i.e., when the excitation probability approaches a critical value of electronically excited atoms [14].

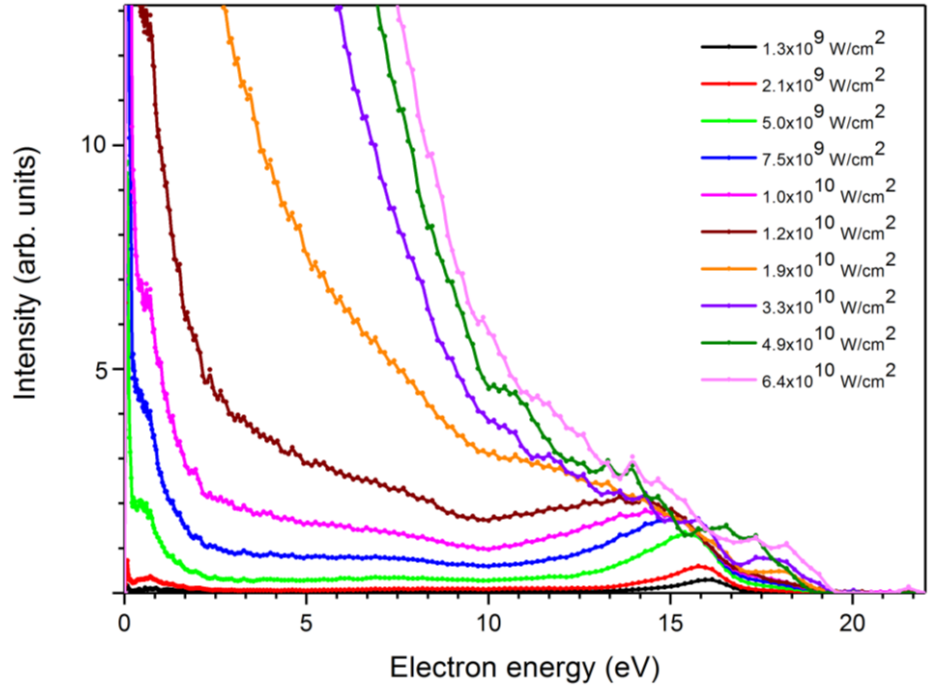

**Figure 14.** Measured electron spectra of extremely large He nanodroplets ($N \sim 1\,000\,000$ atoms) irradiated at hv=21.5 eV and low power density.

## V. SUMMARY AND CONCLUSIONS

The ionization dynamics of He nanodroplets irradiated by intense femtosecond extreme ultraviolet pulses in the range of power densities $10^9$ - $10^{12}$W/cm$^2$ and a photon energy of 21.5 eV have been investigated by photoelectron spectroscopy. Our experimental results are interpreted with the aid of Monte Carlo simulations based on a simplified model of rate equations including various processes such as multi-step ionization [10, 25], ICD [13], secondary inelastic collisions [10, 14] and desorption of electronically excited atoms from He droplets [29], as well as electronic relaxation processes.

In the case of small He droplets (below 1 000 atoms), resonantly excited He-droplet states first effectively and rapidly decay to low-lying 1s2s $^3$S, 1s2s $^1$S and 1s2p $^1$P atomic excited states by droplet-induced transitions. Subsequently, ICD takes place between pairs of electronically excited states, followed by multi-step ionization.

In the case of medium-sized He nanodroplets (a few few thousand atoms), a pronounced broadening of the electron spectra is observed and the different lines start to overlap due to an increase of the total number of created ICD electrons.

In the case of large He nanodroplets (> 50 000 atoms), inelastic electron scattering starts to play significant role and a cold, dense plasma forms. Furthermore, when more than two electronically excited atoms are involved, the ionization dynamics develops from two-body ICD-type processes to collective autoionization (CAI) and higher-order CAI and/or thermal electron emission dominate.

Our results provide a detailed understanding of how autoionization of droplets proceeds from a low excitation power densities characterized by single sharp electron emission lines to a complex ionization process involving many different processes, which eventually results in a cold dense plasma that emits electrons with broad energy distributions. A logical next step will be to perform time resolved measurements in order to directly access the complete time evolution of this system.

# ACKNOWLEDGMENTS

The authors would like to thank P. Demekhin, K. Gokhberg, L. S. Cederbaum, T. Fennel, and U. Saalmann for inspiring discussions. The support of the FERMI staff and financial support by the Deutsche Forschungsgemeinschaft within projects MO 719/14-1, STI 125/19-1 and MU 2347/12-1 (Priority Programme 1840) are gratefully acknowledged.